\documentclass[journal]{IEEEtran}

\usepackage{graphicx}
\usepackage{latexsym}
\usepackage{times}
\usepackage{helvet}
\usepackage{courier}
\usepackage{caption}
\usepackage{booktabs} 
\usepackage{multirow}   
\usepackage{url}
\usepackage{comment, amssymb}
\IEEEtriggeratref{99}

\def\Underline{\setbox0\hbox\bgroup\let\\\endUnderline}
\def\endUnderline{\vphantom{y}\egroup\smash{\underline{\box0}}\\}
\def\|{\verb|}

\usepackage[varg]{txfonts}
\makeatletter%
\input{ot1txtt.fd}
\makeatother%

\begin{document}

\title{Self Voice Conversion as an Attack against \\ Neural Audio Watermarking}

\author{Yigitcan {\"O}zer, Wanying Ge, Zhe Zhang, Xin Wang, Junichi Yamagishi
\thanks{The authors are with National Institute of Informatics, Chiyoda-ku, Tokyo, Japan (e-mail:\{yiitozer,gewanying,zhe,wangxin,jyamagis\}@nii.ac.jp).}}

\maketitle

\begin{abstract}
Audio watermarking embeds auxiliary information into speech while maintaining speaker identity, linguistic content, and perceptual quality. 
Although recent advances in neural and digital signal processing-based watermarking methods have improved imperceptibility and embedding capacity, robustness is still primarily assessed against conventional distortions such as compression, additive noise, and resampling. 
However, the rise of deep learning-based attacks introduces novel and significant threats to watermark security. 
In this work, we investigate self voice conversion as a universal, content-preserving attack against audio watermarking systems. 
Self voice conversion remaps a speaker's voice to the same identity while altering acoustic characteristics through a voice conversion model. 
We demonstrate that this attack severely degrades the reliability of state-of-the-art watermarking approaches and highlight its implications for the security of modern audio watermarking techniques.
\end{abstract}

\begin{IEEEkeywords}
audio watermarking, self voice conversion, robustness, deepfake
\end{IEEEkeywords}

\section{Introduction}
Recent advances in neural audio generation and transformation technologies have led to significant progress in expressive speech synthesis and high-fidelity voice conversion (VC)\,\cite{ShenEtAl23_NaturalSpeech2_ICLR, ChenEtAl25_VALLE_TASLP, BaasNK23_VoiceConversionNN_Interspeech}.
While these developments support novel creative and commercial applications, they have also introduced growing concerns regarding content authenticity, copyright protection, and speaker-identity misuse\,\cite{WangYamagishi21_SpoofingCountermeas_Interspeech}. 
As a result, audio watermarking has become an important line of defense for origin verification, content authentication, tamper detection, and ownership tracking.

Audio watermarking embeds auxiliary information into a speech signal while preserving perceptual quality and enabling reliable detection or extraction\,\cite{HuaEtAl16_WatermarkingReview_SP}. Historically, the field was dominated by DSP-based techniques that embedded watermarks through bit-level manipulation, echo hiding, spread-spectrum modulation, or frequency-domain modifications\,\cite{CedricEtAl00_DataConcealmentAudio_TENCON, ChoEtAl03_EchoWatermarking_IWDW, ChengEtAl02_SpreadSpectrumWatermarking_ICASSP, HuHsu15_AudioWatermarkingDCT_SignalProcessing}. 
The primary threats to these systems were incidental transmission-channel distortions such as lossy compression, additive noise, and resampling.

With the emergence of deep learning (DL)-based watermarking methods\,\cite{SanRomanEtAl24_AudioSeal_ICML, LiuEtAl24_TimbreWatermarking_NDSS, ZhouEtAl25_WMCodec_ICASSP, LiEtAl25_VoiceMark_Interspeech, ChenEtal23_WavMark_arXiv, LiuEtAl23_DeARWatermarking_AAAI, OReillyEtAl24_MaskmarkNeuralWatermarking_ICASSP, LiuEtAl24_GROOT_ACMMM, SinghTLM24_SilentCipher_Interspeech, LiuEtAl25_XAttnMark_ICML}, robustness evaluations have expanded to include a wider range of transformations. 
However, most of these transformations still originate from transmission channels rather than intentional attacks.
Even when attacker-side transformations are considered, they are typically limited to low-tech manipulations such as pitch shifting, time stretching, or simple filtering—operations that do not reflect the capabilities of a motivated adversary.
Consequently, the robustness of current watermarking techniques is systematically overestimated. 
Recent benchmarking studies\,\cite{LiuEtAl24_AudioMarkBench_NeurIPS, OezerChoiEtAl25_RAWBench_Interspeech, WuEtAl25_DeepfakeSpeechDetection_Interspeech, GeWY25_SpeakerIDManipulationWM_ICLR, KovacevicEtAl25_DeepMarkBenchmark_ICLR, OReillyEtAl25_DeepAudioWatermarks_ICLR, WenEtAl25_SoK_AudioWatermarking} reveal that modern neural transformations can severely degrade watermark detectability while maintaining perceptual quality.
Although some approaches reported success against these kinds of attacks, such as neural compression-driven\,\cite{ZhouEtAl25_WMCodec_ICASSP, LiuEtAl25_XAttnMark_ICML, JuvelaWang25_CodecAugmentationWatermarking_ICASSP} and voice cloning-based methods\,\cite{LiEtAl25_VoiceMark_Interspeech}, %
these defense methods still fall short of addressing realistic, intentional, and advanced malicious attacks.

Among such attacks, VC represents a particularly serious and underexplored threat.
VC models can render speech with altered acoustic characteristics while preserving the linguistic content\,\cite{SismanEtAl20_VCOverview_TASLP}, and their increasing accessibility has motivated a parallel line of work on defending or tracing speaker identity under such adversarial attacks\,\cite{DengEtAl23_CatchYouIcan_USENIXSec, RenEtAl23_SpeakerTraceability_ACMMM}.
These methods do not propose watermarking schemes, but instead aim to maintain speaker attribution in the presence of content-preserving yet acoustically transformative models.

\begin{figure*}[t]
    \centering
    \includegraphics[width=0.9\textwidth]{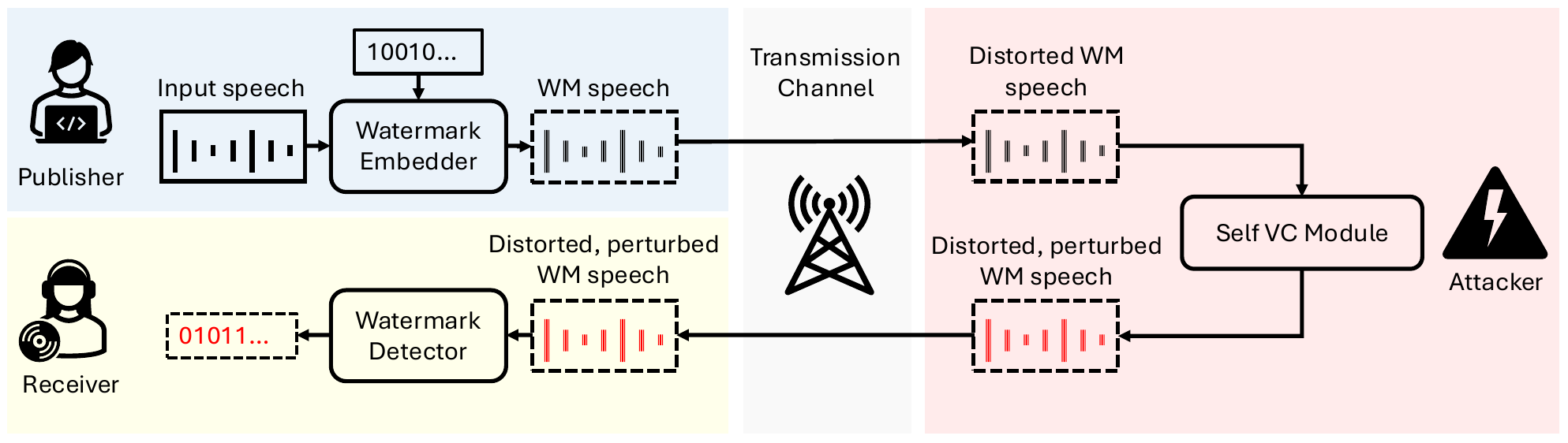} 
    \caption{Illustration of the end-to-end watermarking pipeline under the proposed self voice conversion (VC) attack. The embedder inserts a multi-bit watermark into bona fide speech, which may first undergo transmission channel distortions before reaching the attacker. The attacker applies self VC to regenerate the signal using the same speaker identity, thereby perturbing or suppressing the embedded watermark. Additional transmission channel distortions may also occur after the attacker’s operation, compounding the degradation before the regenerated audio reaches the watermark (WM) detector.}
    \vspace{-0.36cm}
    \label{fig:teaser}
\end{figure*}

In this paper, we address the following critical research question: \emph{Do watermarking systems remain robust when the signal is passed through a VC model?} 
In particular, we investigate \emph{self VC} as a novel and practical attack scenario against audio watermarking systems. 
Self VC remaps a speaker's voice back to the \emph{same} identity, preserving the linguistic content and speaker characteristics while subtly altering the acoustic representation. 
Since the attack does not manipulate the speaker identity or utterance content, it remains imperceptible in many usage contexts; however, it can still significantly disrupt embedded watermarks.

Figure~\ref{fig:teaser} illustrates the end-to-end publisher--attacker--receiver pipeline, where self VC serves as the adversarial attack between watermark embedding and detection. 
This setup reflects realistic deployment conditions in which the attacker acts as an intermediate processing stage without access to the watermarking system's internal details. 
Watermarked speech may also undergo transmission-channel distortions before and after the attacker, further compounding degradation.
Within this framework, we show that self VC poses a severe threat to several state-of-the-art watermarking systems.
Our evaluation covers a classical DCT-based method\,\cite{HuHsu15_AudioWatermarkingDCT_SignalProcessing} and widely used DL-based approaches including AudioSeal\,\cite{SanRomanEtAl24_AudioSeal_ICML}, Timbre\,\cite{LiuEtAl24_TimbreWatermarking_NDSS}, WMCodec\,\cite{ZhouEtAl25_WMCodec_ICASSP}, and VoiceMark\,\cite{LiEtAl25_VoiceMark_Interspeech}\footnote{Audio examples illustrating the effect of self VC on each watermarking system are available at \url{http://yiitozer.github.io/selfvcwm/}.}

\newpage
The remainder of this paper is structured as follows.
Section~\ref{sec:related_work} revisits the specific watermarking models employed in our study and also reviews prior work on audio watermarking attacks.
Section~\ref{sec:proposed_method} introduces the proposed self VC-based attack framework for neural audio watermarking.
Section~\ref{sec:evaluation} presents the empirical results and Section~\ref{sec:discussion} highlights key observations.
Finally, Section~\ref{sec:conclusion} concludes the paper and outlines future directions.

\section{Related Work}
\label{sec:related_work}
This section briefly presents the watermarking algorithms used as proactive defenses in Section~\ref{sec:wm_models}, and then reviews prior work on both transmission channel distortions and attacker-side neural transformations in Section~\ref{sec:attack_methods}.

\subsection{Audio Watermarking Approaches}
\label{sec:wm_models}
We consider five representative audio watermarking approaches that serve as proactive defenses against synthetic speech and voice manipulation.

The first approach is a classical DSP-based blind audio watermarking scheme in the \textbf{DCT} domain, where watermark bits are embedded by adaptively modulating the norm of low-frequency DCT coefficient vectors\,\cite{HuHsu15_AudioWatermarkingDCT_SignalProcessing}.
The embedding strength is constrained by an auditory masking model, and energy compensation with transition smoothing is applied to preserve perceptual transparency and ensure robust watermark recovery.

The second approach is \textbf{AudioSeal}\,\cite{SanRomanEtAl24_AudioSeal_ICML}, which jointly trains a generator to embed an additive watermark signal and a detector to localize watermark presence at the sample level.
AudioSeal operates directly in the waveform domain and its robustness is encouraged during training through the use of an EnCodec-based neural audio codec\,\cite{DefossezCSA23_Encodec_TMLR} and distortion augmentations, including time modifications, filtering, audio effects, and compression.

The third approach is \textbf{Timbre}\,\cite{LiuEtAl24_TimbreWatermarking_NDSS}, a DL-based watermarking method that embeds watermark information in the frequency domain, which is inherently robust to common signal processing operations.
Its training pipeline incorporates iSTFT, normalization, and waveform reconstruction to improve robustness.
Recent benchmark studies report that Timbre achieves stronger robustness compared to other baselines; however, this comes at the cost of reduced imperceptibility\,\cite{LiuEtAl24_AudioMarkBench_NeurIPS, OezerChoiEtAl25_RAWBench_Interspeech}.

The fourth approach is \textbf{WMCodec}\,\cite{ZhouEtAl25_WMCodec_ICASSP}, which introduces an iterative Attention Imprint Unit (AIU) to enhance the fusion of watermark and speech representations.
AIU leverages cross attention to induce a shared latent space that is claimed to enable deep and persistent watermark embedding.

The fifth approach is \textbf{VoiceMark}\,\cite{LiEtAl25_VoiceMark_Interspeech}, a speaker-specific latent watermarking framework designed to be robust against zero-shot voice cloning attacks.
It builds on a pretrained residual vector quantization (RVQ)-based encoder--decoder architecture derived from SpeechTokenizer\,\cite{ZhangEtAl24_SpeechTokenizer_ICLR}, where the first VQ layer captures linguistic content via knowledge distillation from a pretrained HuBERT model\,\cite{HsuBTLSM21_HuBERT_TASLP}, while the remaining layers represent speaker and acoustic attributes.
Robustness is enhanced through augmentation strategies that explicitly simulate voice cloning.

Together, these methods span classical DSP and modern DL-based watermarking paradigms, providing a diverse set of proactive defenses for evaluating the robustness of neural audio watermarking under self VC attacks.

\begin{figure*}[t]
    \centering
    \includegraphics[width=0.9\textwidth]{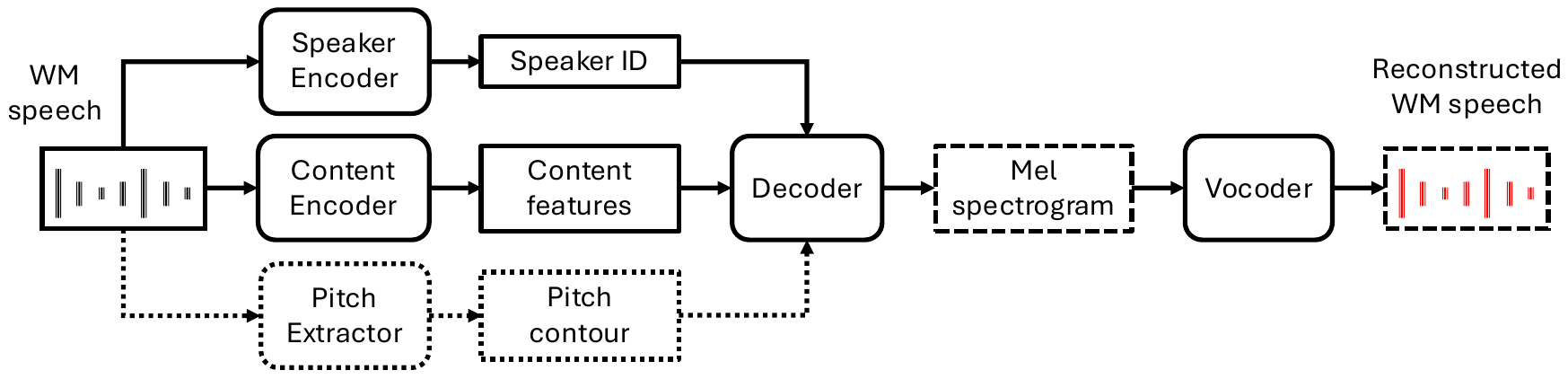} 
    \caption{Overview of the proposed self VC pipeline. Input watermarked (WM) speech is decomposed into speaker identity, content features, and, optionally, pitch contour (extracted via a pitch tracker, shown with dotted lines). These representations are fused by the decoder to generate a mel spectrogram, which is then synthesized into waveform by a vocoder to reconstruct the WM speech signal.}
    \label{fig:self_vc}
    \vspace{-0.36cm}
\end{figure*}

\subsection{From Channel Distortions to Malicious Neural Attacks}
\label{sec:attack_methods}
In this section, we distinguish between two fundamentally different categories of transformations considered in prior robustness evaluations: (i) transmission channel distortions, which arise naturally during storage, streaming, or communication, and (ii) attacker-side transformations, which represent intentional attempts to remove or disrupt embedded watermarks.
This distinction is essential, as many transformations previously treated as \emph{attacks} do not reflect the capabilities of a malicious adversary.

\subsubsection{Transmission Channel Distortions}
Transmission channels introduce a variety of non-malicious distortions, including traditional lossy compression (MP3, AAC, OGG), additive noise, filtering, and resampling. 
These transformations have long served as standard robustness checks for audio watermarking.
Modern neural audio compression represents a more recent and substantially stronger form of transmission channel distortion.
Neural codecs, such as EnCodec\,\cite{DefossezCSA23_Encodec_TMLR} and Descript Audio Codec (DAC)\,\cite{KumarEtAl23_DAC_NeurIPS}, achieve compression ratios exceeding traditional DSP methods by leveraging DL and VQ. 
Since watermarking typically hides information in \emph{redundant} signal components, e.g.,\,high-frequency details, phase relationships, or subtle noise-like patterns, neural codecs fundamentally threaten the integrity of embedded watermarks. 
At low bitrates, RVQ prioritizes perceptually dominant features such as phonetic content, pitch, and spectral envelopes, while residual components carrying watermark information are treated as noise and discarded.
Recent benchmarks such as AudioMarkBench\,\cite{LiuEtAl24_AudioMarkBench_NeurIPS} and RAW-Bench\,\cite{OezerChoiEtAl25_RAWBench_Interspeech} confirm neural codecs as among the most effective distortions.
In parallel, recent watermarking models such as\,\cite{LiuEtAl25_XAttnMark_ICML, JuvelaWang25_CodecAugmentationWatermarking_ICASSP, ZhouEtAl25_WMCodec_ICASSP} exhibit robustness even under neural-codec-driven distortions.

\subsubsection{Attacker-Side Transformations}
In contrast to transmission channel distortions, attackers may intentionally apply transformations designed to suppress or remove watermark information. 
Prior work has considered only a limited set of such transformations. 
Generative restoration attacks, for example, apply \emph{copy synthesis} via neural vocoders such as HiFiGAN\,\cite{KongKB20_HiFiGAN_SpeechSynthesis_NeurIPS} resynthesizing waveforms without altering linguistic or speaker content.
Since the vocoder is typically trained on clean speech, it tends to reject the out-of-distribution artifacts introduced by watermarking, reconstructing a \emph{clean} signal~\cite{LiuEtAl24_TimbreWatermarking_NDSS, OReillyEtAl24_MaskmarkNeuralWatermarking_ICASSP}.
%

%
%
%

A more recent line of work considers zero-shot voice cloning as an adversarial transformation. 
Recent cloning models can synthesize high quality speech from short prompts without retraining\,\cite{ShenEtAl23_NaturalSpeech2_ICLR, ChenEtAl25_VALLE_TASLP}, thereby bypassing watermarking schemes whose robustness relies on training time exposure.
Operating entirely at inference time, these models can alter content, duration, and speaking rate, often destroying watermark information. 
Li et al.\,\cite{LiEtAl25_VoiceMark_Interspeech} show that most watermarking systems collapse to near-random detection under such attacks. 
However, voice cloning typically alters the linguistic content and utterance structure, making it less suitable as a subtle or content-preserving attack.

VC models represent a more subtle and challenging attack class, as they explicitly aim to preserve linguistic content while modifying the acoustic properties of speech, and in some cases also maintaining speaker-related characteristics.
As watermarking mechanisms often rely on fine-grained acoustic cues rather than semantic information, such content-preserving yet acoustically transformative models can disrupt embedded watermarks without introducing perceptible artifacts.
Wen et al.\,\cite{WenEtAl25_SoK_AudioWatermarking} identify VC as a particularly detrimental attack in their study. 

\section{Proposed Attack Method: Self Voice Conversion (VC)}
\label{sec:proposed_method}
We now introduce self VC as an attack against neural audio watermarking in Section~\ref{sec:background_rationale} and then formalize the corresponding threat model and attacker assumptions in Section~\ref{sec:threat_model}.

\subsection{Background and Attack Rationale}
\label{sec:background_rationale}
Modern neural watermarking systems often assume robustness arises from training-time exposure to signal-processing distortions; however, VC reconstructs speech from disentangled latent factors, altering acoustic cues while preserving content and speaker identity.
Recent VC models commonly rely on self-supervised speech representations (S3R), e.g., \,\cite{HsuBTLSM21_HuBERT_TASLP, ChenEtAl22_WavLM_JSTSP}, to encode linguistic content, generate intermediate representations such as mel spectrograms, and synthesize waveforms using neural vocoders.
This pipeline fundamentally differs from waveform-level distortions, as it modifies speech at the representation level rather than through direct signal manipulation.

Our approach, self VC, constitutes a particularly relevant attack setting, in which the input and output speaker identities are identical.
By decomposing speech into disentangled factors and resynthesizing it without explicit quantization, self VC systems can suppress fine-grained signal variations that do not contribute to perceptual quality or identity preservation.
Conceptually, this process resembles an autoencoder or neural audio codec but operates without discrete quantization.
Figure~\ref{fig:self_vc} illustrates the overall self VC pipeline, in which the watermarked input speech waveform is decomposed into content features, speaker identity, and optionally pitch contour (extracted via a pitch tracker such as CREPE\,\cite{KimSLB18_CREPE_ICASSP}, shown with dotted lines).
These components are fused by a decoder to generate a mel spectrogram, which is then synthesized into waveform using a neural vocoder.

\subsection{Threat Model and Attacker's Motivations}
\label{sec:threat_model}
In this work, we consider a threat model in which an attacker aims to invalidate embedded audio watermarks while preserving all perceptually relevant attributes of the original speech signal. 
The attacker’s primary motivation is to remove or disrupt watermark information without altering the speaker identity or linguistic content, ensuring that the transformed audio remains attributable to the same speaker and semantically unchanged. 

The attacker is assumed to have access to a modern VC system capable of decomposing speech into disentangled latent representations, such as speaker identity, linguistic content, and prosodic characteristics. 
%
The attacker can selectively preserve the factors that remain unchanged while unintentionally or intentionally altering latent components that may carry watermark information.

We further assume that the attacker has sufficient technical expertise to operate advanced VC toolkits and computational environments. 
Additionally, regenerated audio can undergo transmission channel distortions, such as additional noise and compression artifacts, before reaching the receiver.
These distortions compound the difficulty of watermark extraction and may further benefit the attacker by masking subtle watermark cues or introducing confounding artifacts that resemble intentional obfuscation.

To evaluate this threat model, we select two VC methods that align with the attacker's goal of preserving speaker identity, maintain linguistic content, and reconstruct speech through latent decomposition processes that are likely to disrupt watermark embeddings. 
These methods represent realistic and practical tools that a motivated adversary could employ in real-world scenarios:

\subsubsection{kNN-VC: VC with Just Nearest Neighbors\,\cite{BaasNK23_VoiceConversionNN_Interspeech}}
\label{sec:knn-vc}
kNN-VC performs VC by replacing each source frame with its nearest neighbors in the reference utterance within a 
S3R space extracted by a pretrained WavLM model \,\cite{ChenEtAl22_WavLM_JSTSP}, which accounts for the balance of phonetic content with speaker identity and prosody.
For each source frame, the model finds the \emph{$k$ nearest frames} in the reference and takes their mean to produce the converted frame. A HiFiGAN vocoder\,\cite{KongKB20_HiFiGAN_SpeechSynthesis_NeurIPS} maps the resulting S3R-based features to output waveforms.

Although kNN-VC does not enforce explicit disentanglement of speech signals, the nearest-neighbor replacement mechanism implicitly suppresses fine-grained signal variations that are not consistently preserved across neighboring representations.
As a result, watermark-related information embedded at the waveform or feature level may be disrupted during the representation matching and resynthesis stages, while speaker identity and linguistic content remain largely intact.

\subsubsection{Retrieval Based Voice Conversion (RVC)}
Instead of kNN-VC, the attacker may use RVC for the self VC attack.
RVC is an easy-to-use VC framework\footnote{\url{github.com/RVC-Project}} built on a conditional variational autoencoder architecture inspired by\,\cite{KimEtAl21_CVAE_TTS_ICML}.
The system uses HuBERT\,\cite{HsuBTLSM21_HuBERT_TASLP} as its content encoder and incorporates pitch estimation via CREPE\,\cite{KimSLB18_CREPE_ICASSP}, as illustrated with dotted lines in Figure~\ref{fig:self_vc}.
While the original RVC implementation is designed for one-to-one voice conversion, we adapt it for Self VC by finetuning the model with an additional speaker embedding layer derived from ECAPA embeddings\,\cite{DesplanquesTD20_ECAPA-TDNN_Interspeech}.
This modification enables both same-speaker and multi-speaker reconstruction while preserving the disentanglement properties that make RVC suitable for watermark-removal attacks.

\begin{table}[t!]
\caption{Evaluation of ground-truth (GT) and VC models. 
}
\centering
\begin{tabular}{lccc}
\toprule
\textbf{Condition} & \textbf{Speaker Similarity $\uparrow$} & \textbf{WER $\downarrow$}  & \textbf{UTMOS\,\cite{SaekiEtAl22_UTMOS_Interspeech} $\uparrow$} \\
\midrule
GT& 1.0   & 0.114 & 4.152 \\
\midrule
kNN-VC      & 0.857 & 0.115 & 3.941  \\
RVC         & 0.748 & 0.120 & 4.190  \\
\bottomrule
\end{tabular}
\label{tab:self_vc}
\vspace{-0.36cm}

\end{table}

\begin{table*}[t]
\caption{Performance comparison across different attack methods. All values show attacker performance values in bitwise accuracy.}

\centering
\begin{tabular}{lllccccc}
\toprule
\multirow{3}{*}{\textbf{Attack Type}} & \multirow{3}{*}{\textbf{Attack Name}} & \multirow{3}{*}{\shortstack{\textbf{Trans.}\\\textbf{Channel}}} &
\multicolumn{5}{c}{\textbf{Watermarking Algorithm}} \\
\cmidrule(lr){4-8}
 & & & \textbf{DCT\,\cite{HuHsu15_AudioWatermarkingDCT_SignalProcessing}} &
   \textbf{AudioSeal\,\cite{SanRomanEtAl24_AudioSeal_ICML}} &
   \textbf{Timbre\,\cite{LiuEtAl24_TimbreWatermarking_NDSS}} &
   \textbf{WMCodec\,\cite{ZhouEtAl25_WMCodec_ICASSP}} &
   \textbf{VoiceMark\,\cite{LiEtAl25_VoiceMark_Interspeech}} \\
\midrule

\multirow{3}{*}{\textbf{No attack}} & \multirow{3}{*}{-} & $\times$ & 0.148 & 0.000 & 0.000 & 0.000 & 0.015 \\

\cmidrule{3-8}
& & $\checkmark$ & 0.405 & 0.032 & 0.045 & 0.034 & 0.216 \\

\midrule
\multirow{9}{*}{\textbf{Baseline Attack: Vocoder}}
&
HiFiGAN\,\cite{KongKB20_HiFiGAN_SpeechSynthesis_NeurIPS}
& \multirow{4}{*}{$\times$}
        & 0.499 & 0.482 & 0.006 & 0.527 & 0.017 \\
& HNSincNSF\,\cite{WangYamagishi19_HNSincNSF_SSW} &
        & 0.497 & 0.490 & 0.194 & 0.521 & 0.030 \\
& HNSincNSFHiFi\,\cite{TomashenkoEtAl22_VoicePrivacy_arXiv} &
        & 0.499 & 0.490 & 0.193 & 0.519 & 0.029 \\
& WaveGlow\,\cite{PrengerVC19_WaveGlow_ICASSP} &
        & 0.497 & 0.493 & 0.025 & 0.539 & 0.055 \\

\cmidrule{2-8}
& 
HiFiGAN & \multirow{4}{*}{$\checkmark$} 
                & 0.499 & 0.495 & 0.186 & 0.500 & 0.333 \\
& HNSincNSF     & & 0.499 & 0.497 & 0.344 & 0.506 & 0.323 \\
& HNSincNSFHiFi & & 0.499 & 0.502 & 0.341 & 0.523 & 0.318 \\
& WaveGlow      & & 0.499 & 0.502 & 0.195 & 0.506 & 0.365 \\

\midrule
\multirow{5}{*}{\textbf{Proposed Attack: Self VC}} &  
kNN-VC\,\cite{BaasNK23_VoiceConversionNN_Interspeech} & \multirow{2}{*}{$\times$} 
        & 0.496 & 0.498 & 0.502 & 0.539 & 0.496 \\
& RVC &
        & 0.490 & 0.499 & 0.490 & 0.501 & 0.498 \\

\cmidrule{2-8}
& kNN-VC & \multirow{2}{*}{$\checkmark$}  & 0.500 & 0.496 & 0.498 & 0.525 & 0.504 \\
& RVC &   & 0.500 & 0.493 & 0.498 & 0.507 & 0.494 \\

\bottomrule
\end{tabular}
\vspace{-0.36cm}

\label{tab:attack_results}
\end{table*}

\section{Evaluation}
\label{sec:evaluation}
We conduct the evaluation on the \emph{test-clean} split of LibriTTS\,\cite{ZenEtAl19_LibriTTSCorpus_Interspeech}.
We first assess whether the proposed self VC models preserve speaker identity and linguistic content, consistent with the threat model in Section~\ref{sec:threat_model}.
Speaker similarity is measured using the cosine similarity between
ECAPA speaker embeddings\,\cite{DesplanquesTD20_ECAPA-TDNN_Interspeech} extracted from reference and converted utterances.
Content preservation is evaluated using the Whisper-L automatic speech recognition (ASR) model\,\cite{RadfordEtAl23_Whisper_ICML}, where word error rates (WER) are computed against the ground-truth (GT) text.
Finally, we assess perceptual speech quality using UTMOS\,\cite{SaekiEtAl22_UTMOS_Interspeech}, a neural mean opinion score (MOS) predictor commonly used in TTS and VC evaluation.

Table~\ref{tab:self_vc} reports the evaluation results for GT speech and the two self VC models.
As expected, GT achieves perfect speaker similarity and the lowest WER, serving as an upper bound.
Notably, both self VC models attain WER values that are close to those obtained when GT speech is directly processed by the ASR model, indicating strong preservation of linguistic content.
Similarly, UTMOS scores for self VC outputs are comparable to those of GT recordings, suggesting that perceptual speech quality is largely maintained.
Between the two self VC systems, kNN-VC better preserves speaker identity and linguistic content, achieving higher speaker similarity ($0.857$) and lower WER ($0.115$) than RVC ($0.748$ and $0.120$, respectively).
In contrast, RVC attains slightly higher perceptual quality, with an UTMOS of $4.190$ compared to $3.941$ for kNN-VC.
Overall, these results indicate that self VC can preserve speaker identity, intelligibility, and perceptual quality to a convincing extent, while still providing a realistic and effective attack setting for evaluating watermark robustness.

To model realistic transmission channel distortions, we apply random compound attacks that combine background noise, Gaussian noise, resampling, and lossy compression.
%
Background noise sampled from the DEMAND dataset\,\cite{ThiemannIV13_DEMAND_PMA} and additive Gaussian noise are randomly mixed at signal-to-noise ratios between $10$ and $30~\mathrm{dB}$.
In addition, lossy compression artifacts are introduced using common audio codecs (AAC, MP3, and Vorbis) at bitrates ranging from $64$ to $192~\mathrm{kbps}$, reflecting typical web and streaming pipelines.
For each utterance, the distortion types and parameters are randomly selected but fixed using a global random seed to ensure reproducibility across attack scenarios.

Table~\ref{tab:attack_results} reports attacker performance as 
$1$ - bitwise extraction accuracy, where lower values indicate stronger watermark extraction and higher values indicate weaker extraction. 
In this scale, values near $0$ correspond to near-perfect extraction, whereas values near $0.5$ indicate extraction close to random guessing, i.e.,\,an effective watermark failure.
Under the \textbf{no-attack} condition, most modern approaches exhibit near-zero error (AudioSeal $0.000$, Timbre $0.000$, WMCodec $0.000$, VoiceMark $0.015$), confirming reliable extraction on clean speech. 
The classical DCT baseline is less reliable even without attack ($0.148$), indicating a higher intrinsic bit error rate relative to the DL-based approaches.

Adding \textbf{transmission channel distortions} increases error rates across all methods, with a substantial increase for DCT ($0.148$ to $0.405$), and particularly visible degradation for VoiceMark (from $0.015$ to $0.216$).
Other DL-based methods remain relatively robust but no longer near-perfect (AudioSeal $0.032$, Timbre $0.045$, WMCodec $0.034$), suggesting that conventional channel effects already erode bitwise recovery.

For the \textbf{baseline vocoder attacks} (without channel distortions), attacker performance rises sharply for several systems, often reaching the near-random regime.
DCT and AudioSeal are driven to approximately $0.49-0.50$ error, and
WMCodec shows a similar range, with an attacker performance around $0.52-0.54$. 
Timbre shows a more mixed pattern, ranging from a low error for HiFiGAN ($0.006$) to higher values for other vocoders (around $0.190$), reflecting its reliance on HiFiGAN as the native vocoder. 
VoiceMark is comparatively resilient to vocoding-only attacks in these runs ($0.017-0.055$), remaining far from the random-guessing region.

The strongest and most consistent degradation arises under the proposed \textbf{self VC attacks}. 
Both kNN-VC and RVC push \emph{all} evaluated watermarking methods toward the random-guessing regime, with values clustered around $0.5$. 
This includes Timbre ($0.502$ under kNN-VC and $0.490$ under RVC) and VoiceMark ($0.496$ and $0.498$), where self VC is markedly more harmful than vocoding-only attacks. 
This failure occurs while the transformed audio remains intelligible and natural, reinforcing that self VC is a practical, content-preserving threat rather than a destructive distortion.
Overall, these results show that self VC acts as a universal attack that drives bitwise extraction to near-random performance while preserving perceptual quality and intelligibility.

\section{Discussion}
\label{sec:discussion}
The results indicate that vocoding alone is often insufficient to remove embedded watermarks, whereas self VC leads to a complete collapse of bitwise extraction performance.
This difference can be attributed to the role of intermediate representations. 
Neural vocoders reconstruct waveforms from acoustic features such as mel spectrograms, possibly allowing any watermark components that survive feature extraction to propagate through resynthesis.
%
%
%
In contrast, self VC typically operates on \emph{disentangled} representations, where linguistic content and speaker characteristics are encoded separately before being resynthesized. 
This refactorization fundamentally alters the acoustic realization of speech, effectively discarding watermark-bearing cues even when perceptual quality and intelligibility are preserved.

The effectiveness of self VC as an attack is closely tied to the use of large pretrained foundation models, e.g.,\,\cite{HsuBTLSM21_HuBERT_TASLP, ChenEtAl22_WavLM_JSTSP} for content and speaker encoding. 
Across all VC systems, we first verify that converted speech preserves both content and speaker identity, confirming that watermark degradation cannot be attributed to trivial distortion. 
These findings suggest that watermarking methods implicitly rely on acoustic properties that are unstable under modern representation learning pipelines, highlighting a fundamental mismatch between current watermark designs and the abstractions learned by speech foundation models.

Defending against self VC presents fundamental challenges for current watermarking paradigms. 
A straightforward response is to apply deepfake detection methods; however, such approaches are inherently \emph{reactive}: they operate only after the signal has been transformed and do not prevent watermark degradation during intermediate processing. 

In a broader sense, the effectiveness of self VC suggests that watermarking schemes relying on fragile acoustic cues are misaligned with modern speech processing pipelines built on learned representations. 
Robust defenses may require watermark designs that are invariant to content-speaker disentanglement or operating directly within the representation spaces used by foundational models.

\section{Conclusion}
\label{sec:conclusion}
In this paper, we introduced self VC as a novel and practical attack method against neural audio watermarking techniques. 
By leveraging VC models that decompose and reconstruct speech while preserving speaker identity and linguistic content, self VC enables attackers to regenerate audio that suppresses watermark information. 
Our experiments demonstrate that this reconstruction-based attack can significantly degrade watermark detection performance, often exceeding the effectiveness of conventional vocoder attacks. 
These findings highlight the need for watermarking techniques that remain robust under latent-space manipulations and self-reconstruction processes, which are increasingly accessible through modern generative speech models.

\section{Acknowledgements}
This work was supported by JSPS MEXT KAKENHI Grant (24H00732); due to potential security implications, source code and model checkpoints will not be publicly released.

\bibliographystyle{IEEEtran}
\bibliography{bibliography.bib}
\end{document}